# Quasi-van der Waals Epitaxial Growth of γ'-GaSe Nanometer-Thick Films on GaAs(111)B Substrates

Mingyu Yu,[†] Sahani Amaya Iddawela,[‡] Jiayang Wang,[§] Maria Hilse,[§,‖,⊥] Jessica L. Thompson,[‡] Danielle Reifsnyder Hickey,[‡,§,‖] Susan B. Sinnott,[‡,§,‖,#,¶] Stephanie Law[*,§,‖,⊥,¶]

AUTHOR ADDRESS

[†]Department of Materials Science and Engineering, University of Delaware, 201 Dupont Hall, 127 The Green, Newark, DE 19716 USA

[‡]Department of Chemistry, The Pennsylvania State University, University Park, PA 16802 USA

[§]Department of Materials Science and Engineering, The Pennsylvania State University, University Park, PA 16802 USA

[‖]Materials Research Institute, The Pennsylvania State University, University Park, PA 16802, USA

[⊥]2D Crystal Consortium Materials Innovation Platform, The Pennsylvania State University, University Park, PA 16802, USA

[#]Institute for Computational and Data Science, The Pennsylvania State University, University Park, PA 16802, USA

¶Penn State Institute of Energy and the Environment, The Pennsylvania State University, University Park, PA 16802 USA

Corresponding Email: sal6149@psu.edu

ABSTRACT: GaSe is an important member of the post-transition metal chalcogenide family and is an emerging two-dimensional (2D) semiconductor material. Because it is a van der Waals material, it can be fabricated into atomic-scale ultrathin films, making it suitable for the preparation of compact, heterostructure devices. In addition, GaSe possesses unusual optical and electronic properties, such as a shift from an indirect-bandgap single-layer film to a direct-bandgap bulk material, rare intrinsic p-type conduction, and nonlinear optical behaviors. These properties make GaSe an appealing candidate for the fabrication of field-effect transistors, photodetectors, and photovoltaics. However, the wafer-scale production of pure GaSe single crystal thin films remains challenging. This study develops an approach for the direct growth of nanometer-thick GaSe films on GaAs substrates using molecular beam epitaxy. It yields smooth thin GaSe films with the rare γ'-polymorph. We analyze the formation mechanism of γ'-GaSe using density functional theory and speculate that it is stabilized by Ga vacancies since the formation enthalpy of γ'-GaSe tends to become lower than that of other polymorphs when the Ga vacancy concentration increases. Finally, we investigate the growth conditions of GaSe, providing valuable insights for exploring 2D/3D quasi-van der Waals epitaxial growth.

In recent decades, layered chalcogenides have garnered significant attention as advanced members in the field of two-dimensional (2D) semiconductors due to their wide range of stoichiometries and stacking sequences, broadly tunable band gaps,[1-9] and versatile optical properties.[10-14] Within this large family, GaSe is of interest for its potential in optical and optoelectronics applications. GaSe differs from most other 2D semiconductors in that the bandgap of GaSe changes from an indirect transition of 3.3 eV to a direct transition of 2.1 eV as the film thickness increases from one layer to bulk.[14-16] Moreover, in a single layer of GaSe, the energy gap of the direct transition is only 0.092 eV higher than that of the indirect transition, making GaSe readily convertible into a direct-bandgap material through external stimuli even at atomically thin thicknesses. GaSe also exhibits intrinsic p-type conductivity, which is rare among 2D materials.[10,17] Additionally, GaSe has nonlinear optical characteristics in the infrared spectrum[11,12] and high transparency from 650 nm to 18000 nm.[13,14] These optical and electronic properties make GaSe an outstanding material for fabricating field-effect transistors,[18] photodetectors,[19-22] and photovoltaic devices.[23] Finally, the relatively low growth temperature required for high-quality GaSe enables its integration into the back-end-of-line semiconductor processes, thus providing a platform for the scalable production of GaSe devices.

However, obstacles persist in the widespread adoption of GaSe semiconductors, with wafer-scale and nanometer-thick synthesis being major challenges. To date, exfoliation remains the predominant approach for synthesizing GaSe thin films, but is plagued by impurities, imprecise film thickness control, and mass production challenges.[24-26] Hence, alternative synthesis methods, including chemical vapor deposition,[27] pulsed laser deposition,[28] and molecular beam epitaxy (MBE),[16,29-34] are under investigation. MBE presents distinct advantages in growing high-quality GaSe films due to its high-purity environment, precise thickness control, and wafer-scale growth platform. Also, MBE excels in accurate regulation of the flux ratio and growth temperature, which is essential to attain single-phase and single-polymorph GaSe films. Ga-Se compounds have multiple stable phases, such as GaSe and $Ga_2Se_3$, and multiple polymorphs which have similar formation energies.[31] A single layer of GaSe consists of four atomic planes that are covalently bonded in the Se-Ga-Ga-Se sequence, called a tetralayer (TL). The weak interlayer van der Waals (vdW) bonds enable a variety of possible stacking orders and polytypes including ε- (2R), β- (2H), δ- (4H), and γ- (3R), all of which have a non-centrosymmetric TL and a point group of $D_{3h}$. A rare polymorph, γ'-GaSe, was experimentally observed by Grzonka *et al*. in 2021.[29] It differs from other GaSe polymorphs in that it has a centrosymmetric TL with a point group of $D_{3d}$ that is stacked in the same configuration as γ-GaSe, as shown in the atomic models in Figure 1.

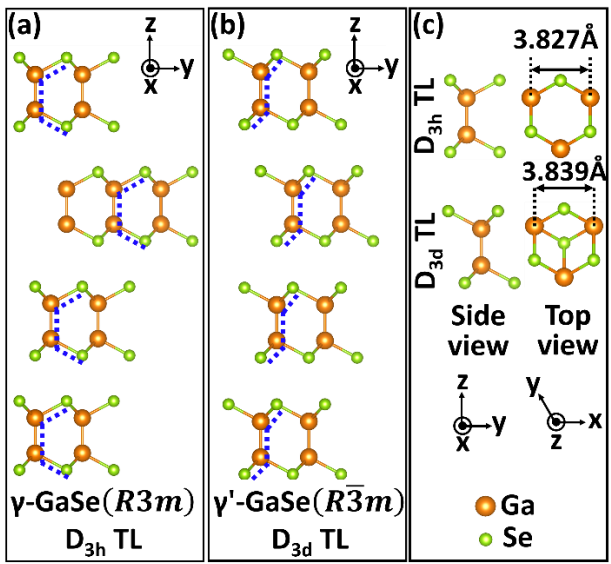

Figure 1. Schematic models of (a) γ-GaSe, (b) γ'-GaSe, (c) side (left) and top (right) views of two GaSe TL polymorphs, $D_{3h}$ (axial symmetry, top) and $D_{3d}$ (central symmetry, bottom). The blue dashed lines highlight the stacking configuration. The thickness of GaSe TL is about 8 Å. The in-plane lattice parameters of $D_{3h}$ and $D_{3d}$ GaSe TL are 3.827 Å and 3.839 Å, respectively.

Extensive research has been done on the synthesis of GaSe on passivated substrates like sapphire[29,30-32,34] and mica,[16] but the weak film/substrate interaction leads to random in-plane alignment of the GaSe crystallite grains, degrading the optical and transport performance. In general, attempts to grow GaSe on 3D semiconductor substrates have not been satisfactory because the large lattice mismatch between the substrate and the film as well as a large number of dangling bonds on the substrate surface make it challenging to obtain wafer-scale flat and continuous GaSe single crystal films. In this paper, we demonstrate the quasi-vdW MBE[35-37] growth of GaSe nanometer-thick films on GaAs(111)B substrates, which is a mature 3D semiconductor material with a hexagonal lattice similar to GaSe and a relatively modest lattice mismatch of −6.4 % (the in-plane lattice constants of GaAs(111) and GaSe(002) along $[0\bar{1}1]$ and $[1\overline{\sqrt{3}0}]$ are 3.998 Å and 3.742 Å, respectively). We systematically studied the growth window for GaSe single crystal films with high structural quality. Of particular interest, we observed the unusual γ'-GaSe polymorph on GaAs(111)B using annular dark field-scanning transmission electron microscopy (ADF-STEM), and analyzed the formation mechanism of γ'-GaSe using density functional theory (DFT) calculations. We found that GaAs(111)B is more suitable than c-plane sapphire (c-sapphire) as a substrate for GaSe growth: the GaSe films grown on GaAs(111)B have a more ordered in-plane alignment and an improved epitaxial relationship between the film and the substrate.

This paper will help advance the wafer-scale production of high-quality γ'-GaSe crystalline films, which are expected to exhibit enhanced optoelectronic properties like second harmonic generation due to the central symmetry of the γ'-GaSe TL.[34] More importantly, we elucidate a clear approach for the direct growth of hybrid 2D/3D heterostructures using MBE technology. The hybrid 2D/3D heterostructure with an atomically sharp interface is expected to inherit the characteristics of both 3D GaAs and 2D GaSe semiconductors, thereby expanding its potential in device applications and laying a solid foundation for the development of integrated quantum photonic devices.

RESULTS AND DISCUSSION

**Study of GaSe Growth Parameters.** The substrate temperature, film growth rate, and atomic flux ratio are three key variables that influence the quality of the MBE-grown films. The nominal growth rate is determined by the scarce element (here Ga flux); however, the actual growth rate is also affected by the growth temperature since high temperatures will cause film re-evaporation. Detailed growth conditions for the samples studied in this paper are summarized in Table 1 and Table S1 (Supporting Information). According to calibration results, Ga fluxes of 1.3 ($\pm$ 0.1) $\times$ $10^{13}$ and 2.3 $\times$ $10^{13}$ atoms cm$^{-2}$ s$^{-1}$ can lead to GaSe growth rates of 0.07 Å s$^{-1}$ and 0.13 Å s$^{-1}$, respectively, when there is no re-evaporation.

Table 1. Growth parameters of GaSe Sample #1 – #8 grown on GaAs(111)B substrates

| Sample | Ga flux [$10^{13}$ atoms cm$^{-2}$ s$^{-1}$] | Se:Ga | Growth temperature [°C] | Growth time [s] |
|---|---|---|---|---|
| #1 | 1.3 | 1.4 | 400 | 2400 |
| #2 | 1.4 | 2.2 | 400 | 2400 |
| #3 | 1.3 | 2.8 | 400 | 2400 |
| #4 | 1.2 | 9.5 | 400 | 2400 |
| #5 | 1.3 | 2.2 | 440 | 2400 |
| #6 | 1.3 | 2.2 | 420 | 2400 |
| #7 | 1.4 | 2.2 | 375 | 2400 |
| #8 | 2.3 | 2.2 | 420 | 2400 |

An oversupply of Se is necessary to compensate for Se re-evaporation from the growth front,[38,39] so it is unsurprising that the GaSe film forms within a broad Se:Ga window of 2.2 to 9.5, as shown in Figure 2a. The 2θ/ω high-resolution X-ray diffraction (HRXRD) scans for Sample #2 – #4 all confirm the formation of GaSe single crystals by detecting three characteristic peaks at 2θ = 11. 1 °, 22.3 °, and 57.7 °, corresponding to the GaSe-(002), (004), and (00$\underline{10}$) planes, respectively. Figure 2b further shows that as the Se:Ga ratio increases, the full-width at half maximum (FWHM) of the GaSe diffraction peaks decreases. This implies that excess Se enhances the quality of GaSe crystals. Unfortunately, the atomic force microscopy (AFM) images in Figure 3 show that this enhanced crystal quality comes at the expense of surface smoothness and film coalescence: the surface morphology of the GaSe film is extremely sensitive to the supply of Se. Se:Ga flux ratios as low as 1.4 failed to generate continuous GaSe crystal films but only produce isolated, spiral, near-triangular flakes (Figure 3a). The triangular spiral morphology is characteristic of 2D chalcogenides.[39-44] The literature[44] on MBE growth of InSe, a material with an almost identical crystal structure to GaSe, has proposed two possible explanations for the spiral growth: 1) unequal growth rates along the zigzag and armchair edges, in combination with the angle of the initial nuclei relative to substrate step edges, can generate spiral dislocation centers for subsequent growth; 2) the metal vacancies created by the Se-rich growth conditions can introduce localized structural distortions, resulting in islands with zigzag edge fronts climbing over neighboring islands rather than merging into 2D layers. We experimentally observed that appropriately increasing the growth temperature can inhibit the spiral growth, which will be discussed in Section 2.1.2, while simply reducing the Se:Ga flux ratio cannot inhibit the generation of spiral centers, as revealed by Figure 3a. We therefore deduce that the former explanation is the dominant factor triggering the spiral growth of GaSe.

Figure 3b illustrates that raising the Se:Ga ratio to 2.2 results in a coalesced, smooth GaSe crystal film composed of the expected triangular domains with multiple spiral centers. A linecut (Figure S1, Supporting Information) taken from the white solid line in Figure 3b shows that the step height between adjacent layers in a spiral is close to the GaSe TL thickness of 8 Å, consistent with the layer-by-layer growth mode of MBE. However, upon slightly raising the flux ratio to 2.8, Figure 3c exhibits a GaSe film with a dramatically increased surface roughness. In addition, the GaSe domains nucleated and grew in a "flower" shape rather than the typical triangular pattern, leading to poorly coalesced island features. The issue worsens when the flux ratio was further increased to 9.5: Figure 3d shows a surface morphology with numerous 3D islands, and the scanning electron microscopy (SEM) image in Figure 4a demonstrates the columnar growth of GaSe crystals. Combining the HRXRD and AFM results, we conclude that changing the flux ratio over the range of 2.2 to 9.5 does not change the film stoichiometry, but excess Se disrupts the initial GaSe nucleation and subsequent coalescence of individual GaSe nuclei. Our hypothesis rests on the fact that the hexagonal GaSe unit cell has two in-plane edges: $[0\bar{1}1]$ and $[1\bar{1}0]$. Each atom on these two edges has two or one dangling bonds, respectively, as depicted in Figure 4b. The resulting faster growth along the $[0\bar{1}1]$ edge yields the characteristic triangular nucleation domains. However, when the Se:Ga flux ratio exceeds the ideal threshold, the growth rate difference between edges no longer regulates the nucleation shape. Instead, excess Se adatoms may cause a reconstruction of the Se dangling bonds on the sidewalls of the GaSe domains, leading to irregular flower-like nucleation and growth habit. In addition, the reconstruction may reduce the number of Se dangling bonds on the sidewalls, causing Ga adatoms to preferentially incorporate on the top of the GaSe domains rather than the sidewalls, resulting in the columnar growth dynamic shown in Figure 4a.[45] Consequently, we need to balance the pros and cons of an excess Se supply: a high Se flux compensates for Se re-evaporation and improves the GaSe crystal quality but reduces the smoothness and coalescence of the GaSe films. This trade-off imposes stringent requirements on the desirable Se:Ga flux ratio for growing high-quality GaSe crystal films, which has been ultimately determined to be between 2.1 and 2.3. A similarly narrow flux ratio window has been reported for the growth of GaSe on sapphire substrates.[32]

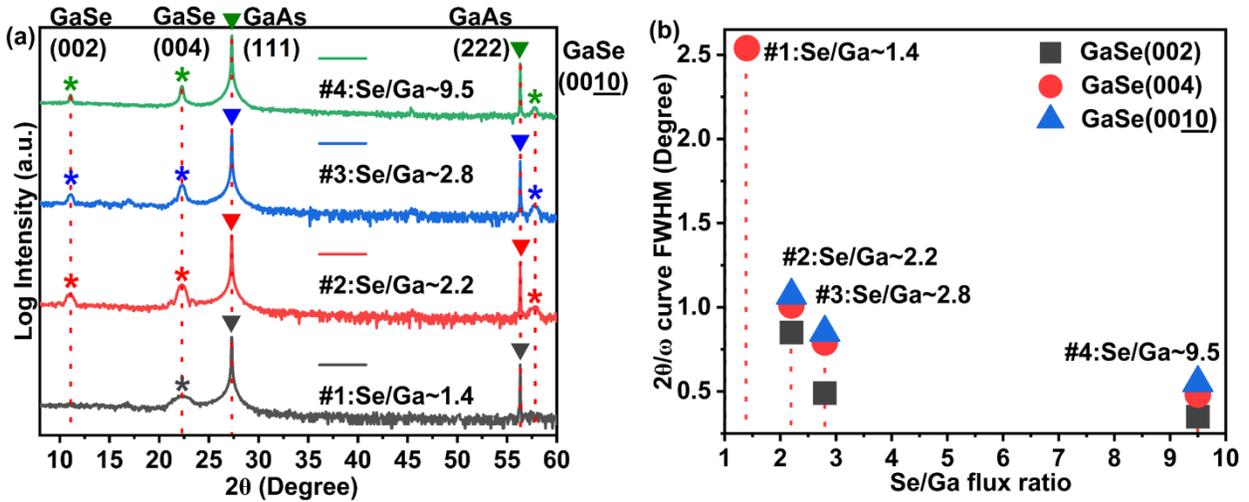

Figure 2. (a) 2θ/ω HRXRD scans of Sample #1 – #4. "*" and "▼" symbols mark the peaks of GaSe and GaAs, respectively. (b) FWHM vs. Se:Ga flux ratio plots for Sample #1 – #4. The FWHM values are from panel (a).

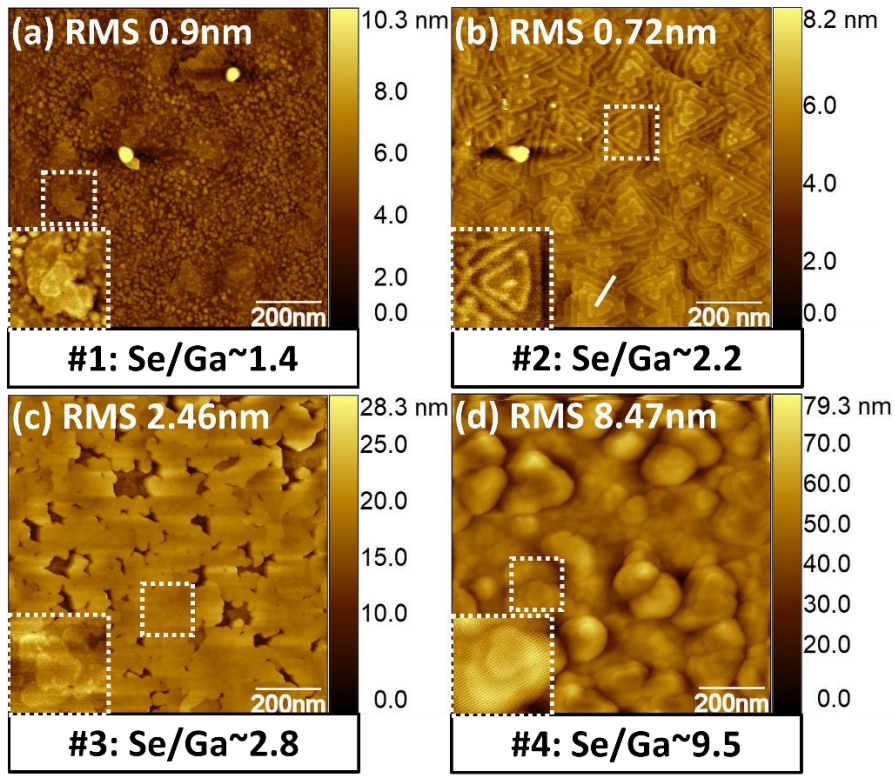

Figure 3. AFM images of Sample (a) #1, (b) #2, (c) #3, and (d) #4. Each AFM image has an inset in the lower left corner that is a 1.9 × magnified view of the area marked by the dashed box (160 nm × 160 nm). RMS indicates the root mean square surface roughness.

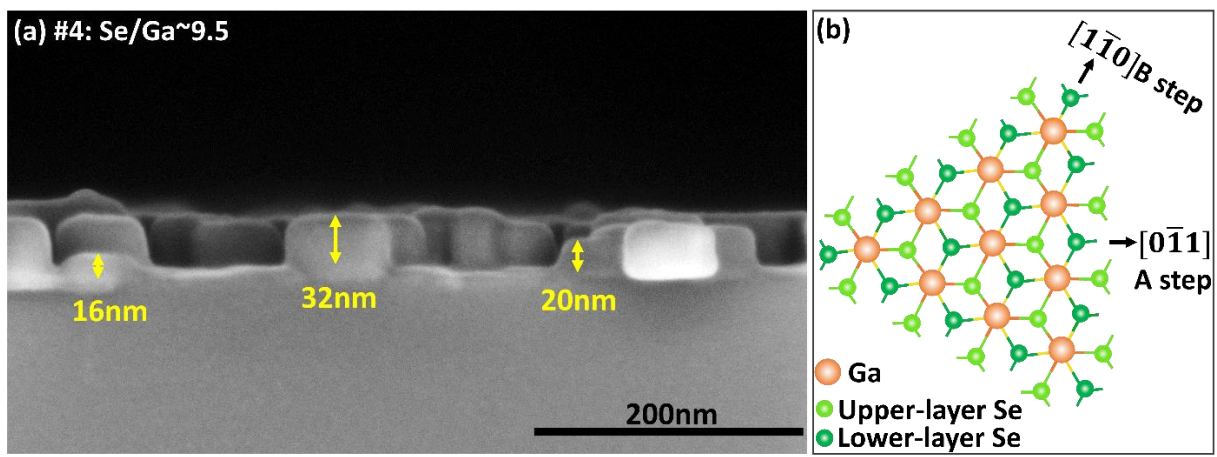

Figure 4. (a) Cross-sectional SEM image of Sample #4. (b) Schematic model of a 2D island of γ'-GaSe, showing type A [0$\bar{1}$1] steps with two dangling bonds per edge atom and type B [1$\bar{1}$0] steps with one dangling bonds per edge atom.

The second important parameter impacting the film morphology is the substrate temperature. In theory, high-temperature growth facilitates large, well-ordered crystallites due to the long adatom diffusion length. However, excessively high temperatures can cause GaSe to decompose and re-evaporate.[46] As exemplified by Sample #5 grown at 440 °C, both the 2θ/ω HRXRD scans (Figure 5a) and the AFM scans (Figure 5b) indicate that GaSe films cannot form at this high substrate temperature. Sample #6, grown at a cooler substrate temperature of 420 °C, comprises GaSe crystallites nucleated in a triangular pattern, as shown in Figure 5c. However, visible cracks are seen on the film which can be attributed to the simultaneous film formation and decomposition/re-evaporation. In contrast, an even lower substrate temperature of 400 °C yielded excellent surface morphology with low roughness and no significant 3D defects, as shown in Figure 3b for Sample #2. Continuing to lower the substrate temperature to 375 °C for Sample #7 did not result in any improvement in the FWHM values given by the 2θ/ω HRXRD scans (Figure 5a) nor in the RMS roughness from AFM (Figure 5d). However, at 375 °C, the GaSe domains nucleated in an irregular pattern rather than the typical triangular pattern (although they somewhat preserved a triangle-like shape), which agrees with the anticipated outcome of less ordered nucleation at lower temperatures due to the reduced adatom mobility.

Interestingly, a relatively high growth temperature (*e.g.*, 420 °C) combined with an appropriately increased Ga flux (while keeping the Se:Ga ratio constant) results in coalesced GaSe films with a significant reduction in screw defects, as evidenced by the AFM image (Figure 5e) of Sample #8. In Section 2.1.1, we determined that unequal growth rates along different edges of the domain combined with the impact of substrate step edges results in the spiral growth of GaSe. High-temperature growth can decrease the density of screw defects by providing additional thermal energy, enabling both fast (Step A) and slow (Step B) growth facets to easily climb over substrate step edges and merge into a 2D layer. It may also promote the re-evaporation of domains that are pinned at a step edge, further reducing the density of spiral dislocations. Another advantage of high-temperature growth is the elimination of misoriented crystallite grains since they more readily re-evaporate at high temperatures, which will be discussed in Section 2.2.1. It is worth noting that a corresponding increase in Ga flux is critical for high-temperature growth, otherwise re-evaporation introduces film cracks and screw dislocations, as shown in the AFM image (Figure 5c) of Sample #6. However, most samples in this study used a growth temperature of 400 °C unless otherwise specified, because the re-evaporation of GaSe at 420 °C makes it challenging to control the actual growth rate. For example, Sample #8 was grown for the same amount of time as Samples #2 and #7 with a larger Ga flux. We would therefore expect a thicker film. However, Sample #8 was only ~10 nm thick, much thinner than Samples #2 and #7 (~16 nm), due to simultaneous deposition and re-evaporation. In addition, we found that GaSe evaporation may result in the formation of GaSe droplets on the film surface: in the Supporting Information, optical microscopy (Figure S2a) and SEM (Figure S2b) detected the presence of circular 3D features; Raman spectra (Figure S2c) and SEM Energy Dispersive Spectroscopy (EDS) mapping (Figure S2d-f) further indicated that the features are composed of GaSe. These features are not reflected in the AFM image (Figure 5e), possibly due to their low density.

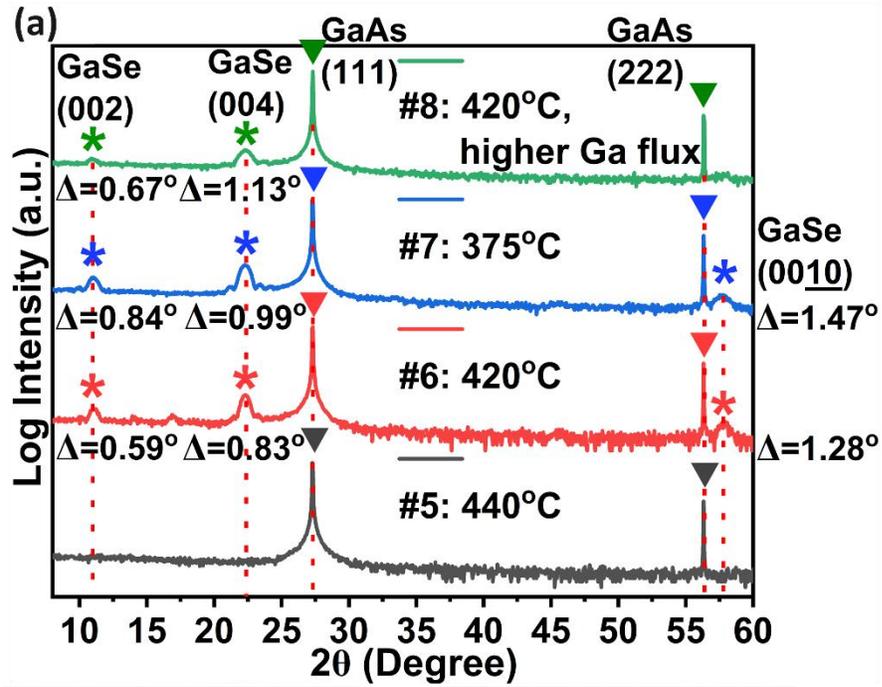
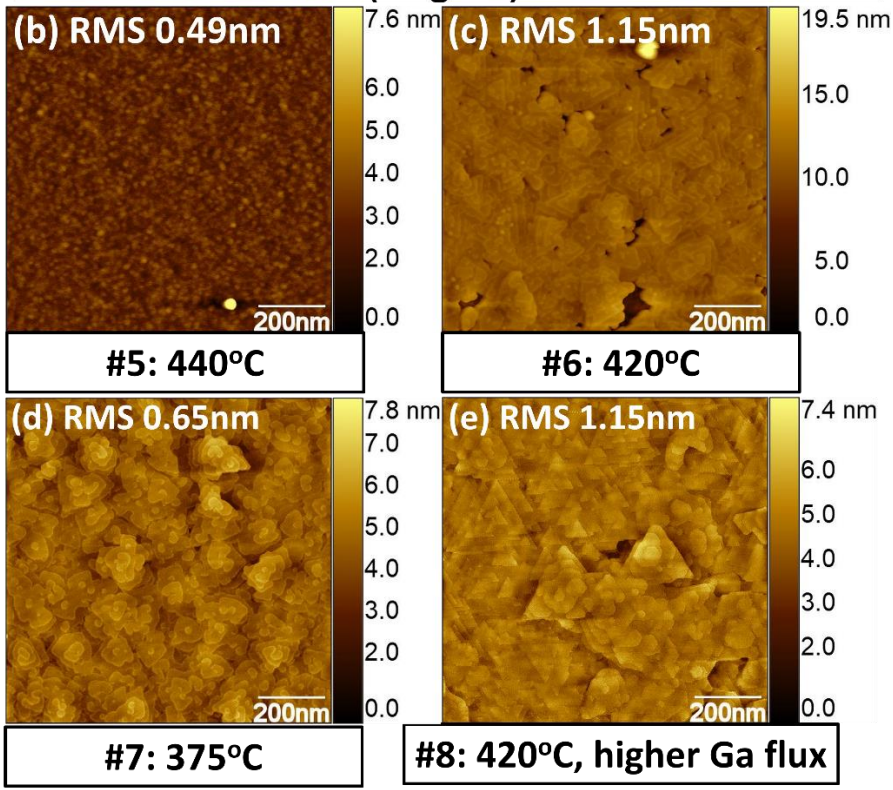

Figure 5. (a) 2θ/ω HRXRD scans of Sample #5 – #8. "*" and "▼" symbols mark the peaks of GaSe and GaAs, respectively. "Δ" indicates the value of FWHM. AFM images of Sample (b) #5, (c) #6, (d) #7, and (e) #8.

In the absence of other factors (*e.g.*, GaSe decomposition/evaporation at high temperatures), the growth rate itself does not cause visible differences in the film morphology, composition, or crystallite size, as evidenced by AFM images (Figure S3, Supporting Information) and 2θ/ω HRXRD scans (Figure S4a, Supporting Information). However, ω scans (Figure 6) reveal that a higher growth rate minimizes defects such as mosaicity, dislocations, and curvature, which can disrupt the parallelism of the atomic planes. The FWHM of the rocking curve narrows as the growth rate increases. The improvement in crystal quality is particularly pronounced when the growth rate is increased from 0.03 Å s$^{-1}$ to 0.05 Å s$^{-1}$. Further growth rate increases only slighμy enhance crystal quality. Considering the constraints on thickness control and Se supply, 0.07 Å s$^{-1}$ was chosen as the appropriate rate for this study, unless otherwise specified.

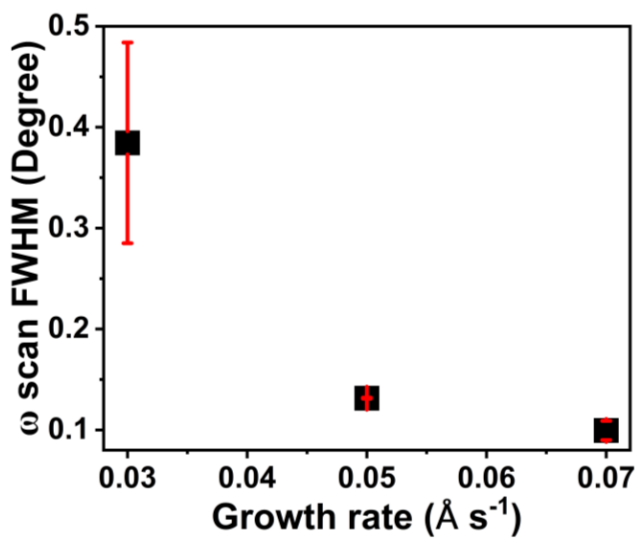

Figure 6. Plot of FWHM vs. growth rate. The FWHM corresponding to each growth rate is the average of the FWHM of the rocking curves shown in Figure S4b, c (Supporting Information). The red vertical lines represent error bars.

**Structural Quality of Optimal GaSe Films.** Through experiments, the optimal set of conditions for growing GaSe on GaAs(111)B were determined to be: a substrate temperature of 400 °C, a Se:Ga flux ratio of 2.2, and a growth rate of 0.07 Å s$^{-1}$. Next, we will analyze the structural quality of the GaSe films obtained under these conditions.

Sample #2 was grown with the optimal parameters. The sharp streaky reflection high-energy electron diffraction (RHEED) patterns (Figure 7a,b) taken during the growth of Sample #2 along the [1$\bar{1}$00] and [11$\bar{2}$0] directions demonstrate that the ratio of streak spacing on the a-plane (11$\bar{2}$0) to that on the m-plane (1$\bar{1}$00) is close to $\sqrt{3}$, indicating the expected 6-fold symmetry of the hexagonal GaSe structure. Here the [1$\bar{1}$00] and [11$\bar{2}$0] directions of GaSe are aligned with the [0$\bar{1}$1] and [2$\bar{1}\bar{1}$] directions of the GaAs(111)B substrate, respectively.[47] The coexistence of a- and m-plane streaks implies that the GaSe film has domains that nucleate in both the [1$\bar{1}$00] and [11$\bar{2}$0] directions. This phenomenon is generally attributed to the similarity in the substrate surface energy along the two directions.[31,48-51]

The in-plane φ scan for GaSe (red curve Figure 7c) shows sharp diffraction peaks every 60 °, which is expected for the hexagonal space group, and there are three peaks exactly overlapping the substrate diffraction peaks (black curve in Figure 7c), indicating a strong epitaxial relationship and excellent alignment between the film and the substrate. The 6-fold symmetry of the φ scan for the GaSe film indicates significant twinning, in which domains are nucleated in opposite directions with equal probability. In addition to the six main peaks, the GaSe film also shows four weak peaks at 23 °, 148 °, 205 °, and 336 °, respectively. The information obtained from the φ scans was visually confirmed in the AFM image (Figure 7d): most of the triangular grains are along the $<0\bar{1}1>$ directions, as indicated by the white arrows, and are well aligned with the major flat of the substrate, indicating a strong epitaxial relationship between the film and substrate. The white arrows point left and right with almost equal probability. The weak peaks in the φ scan are caused by a few GaSe grains which are rotated in plane by ~96 ° relative to the main orientation, as depicted in Figure 7e. These grains are also twinned, as indicated by the blue arrows in Figure 7d. The presence of twin boundaries confirms that the formation energy of GaSe domains in different directions is energetically similar. The main and minor orientations of the GaSe domains coincide with the major and minor flats of the GaAs(111)B wafer, respectively, suggesting that the directional preference in GaSe nucleation is primarily influenced by the substrate. Therefore, it is possible that the use of miscut substrates could bias the film to nucleate in one preferred orientation, reducing or eliminating the twin defects. On the other hand, as mentioned in Section 2.1.2, high-temperature growth can suppress the generation of misoriented GaSe grains that are rotated by 96 ° relative to the main orientation. This is evident by comparing the φ scans of Sample #2 and #8 (Figure 7c, f). This improvement may be due to the fact that misoriented GaSe nuclei are easier to re-evaporate at high temperatures compared to predominately oriented nuclei. Because misoriented nuclei are not preferred at the growth front, they are more likely to form later and at smaller sizes. Thus, high-temperature growth favors the reduction of grain boundaries by promoting larger grain sizes and lowering nucleus the density of nuclei.

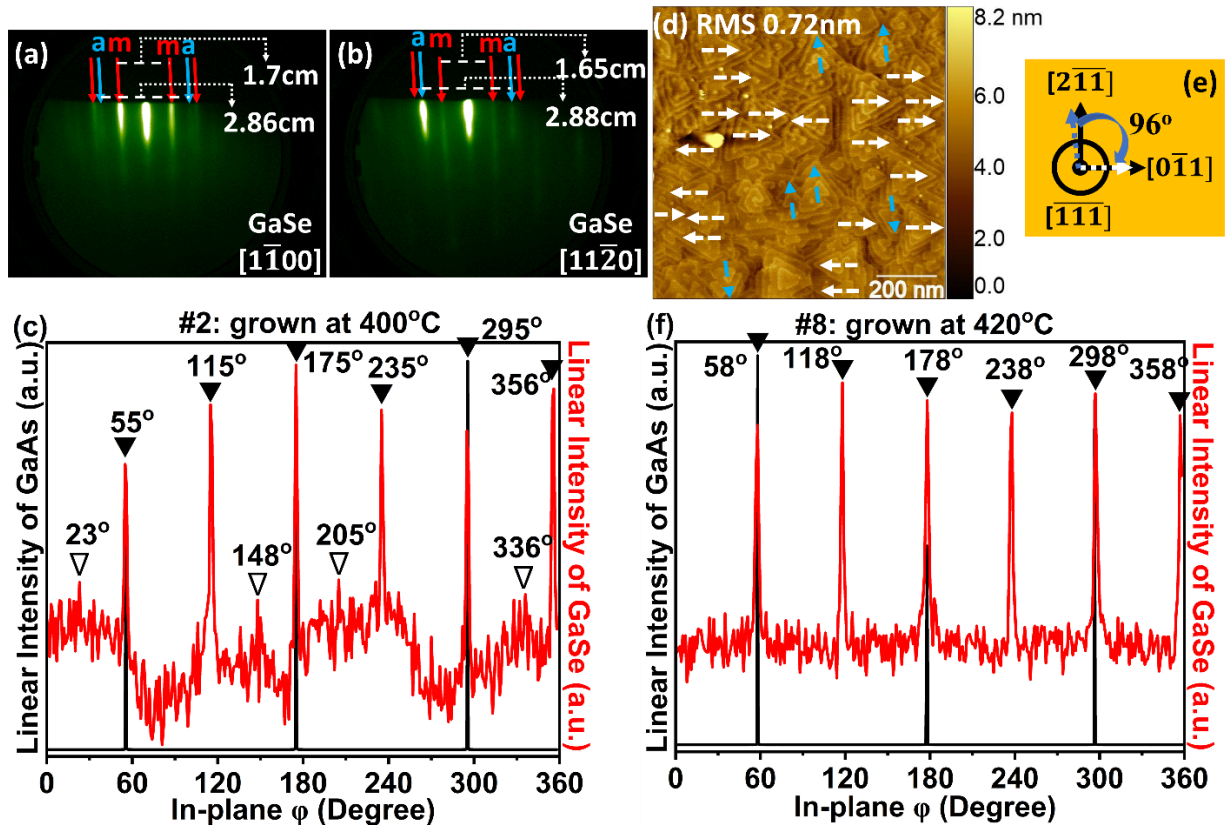

Figure 7. RHEED patterns taken along the (a) [1$\bar{1}$00] and (b) [11$\bar{2}$0] directions during the growth of Sample #2. The m-plane and a-plane denoted by red and blue arrows correspond to the (1$\bar{1}$00) and (11$\bar{2}$0) planes, respectively. The 1.7 cm and 2.86 cm in part (a) (1.65 cm and 2.88 cm in part (b)) are the distances between adjacent m-planes and between adjacent a-planes, respectively. (c) In-plane φ scans of Sample #2. "▼" and "▽" designate the main and minor diffraction peaks of GaSe, respectively. The in-plane scans of the GaSe film and GaAs substrate were around the (103) and (311) planes, respectively. The out-of-plane orientations for GaSe and GaAs were (002) and (111), respectively. (d) AFM image of Sample #2 (the same AFM as Figure 3b). The white and blue dashed arrows highlight the main and minor orientations of GaSe domains, respectively. (e) Schematic diagram depicting the azimuthal angles corresponding to the white and blue dashed arrows in part (b). (f) In-plane φ scans of Sample #8.

A noteworthy finding is that the GaSe film we grow on GaAs(111)B is the rare γ'-polymorph, as determined by ADF-STEM images (Figure 8). Although the contrast between Ga, Se, and As atoms is low due to the small difference in their atomic numbers ($Z_{Ga}$ = 31, $Z_{Se}$ = 34, $Z_{As}$ = 33), we can still see the smooth interface between the GaSe film and GaAs substrate in Figure 8a. The initial GaSe tetralayer (TL) is well-ordered, and this orderliness continues into the subsequent GaSe layers. The high-magnification image (Figure 8b) gives a clearer perspective. As the overlaid atomic sketch indicates, the initial GaSe TL presents an inverted "C" shape, implying that it has the $D_{3h}$ symmetry characteristic of the common GaSe polytypes (β-, ε-, γ-, δ-). However, subsequent GaSe TLs are all "S" shaped, exhibiting a $D_{3d}$ central symmetry, and stacked in a γ-type, thus forming the γ'-GaSe polymorph. γ'-GaSe has seldom been found in layers far from the substrate and usually coexists with other polymorphs, because the formation enthalpies of different polymorphs of GaSe are very close, and the γ'-type is slightly less energetically stable than other polymorphs.[29,34,52] Our DFT calculations confirm this conclusion, as the formation enthalpies obtained for ε-, γ-, and γ'-GaSe are -0.614, -0.614, and -0.611 eV atom$^{-1}$, respectively. We speculate that the formation of γ'-GaSe on GaAs substrates may be attributed to the presence of Ga vacancy defects, which is common during the growth of Ga-chalcogenides.[53] The DFT simulation results in Figure 9a,b demonstrate the validity of this hypothesis: when the Ga vacancy concentration exceeds 0.69 %, 0.71 %, and 0.87 % in single-layer, bulk γ-, and bulk ε-GaSe crystals, respectively, the formation enthalpy of γ'-GaSe ($D_{3d}$ TL) becomes lower than that of other polymorphs, suggesting a tendency to form γ'-GaSe. We further analyzed the cause of this phenomenon from the crystal configuration. After introducing a Ga vacancy and performing DFT relaxation, the Ga atom closest to the vacancy site will move toward the center of the layer and bond with the nearest six Se atoms. Simultaneously, these Se atoms move closer to the central Ga atom. In this case, if the ε/γ-GaSe polymorph is formed, the central Ga atom forms a trigonal prismatic polyhedron with six neighboring Se atoms, where the Ga-Se bond length is ~2.834 Å. In contrast, in γ'-GaSe, an octahedron is formed with Ga-Se bond length of ~2.776 Å, which is closer to the Ga-Se equilibrium bond length of 2.504 Å. Therefore, the presence of Ga vacancies makes the formation of Ga-Se octahedron energetically favorable, leading to the establishment of inversion symmetry in the GaSe layer, as shown in Figure 9c. We also examined the addition of Se vacancies, but an increase in Se vacancy concentration instead promotes the stability of ε-GaSe, as indicated in Figure S5 (Supporting Information). Finally, we attribute the formation of the initial $D_{3h}$ symmetry at the interface to the interference of the substrate, despite the effective passivation of dangling bonds on the substrate surface by Se atoms, as depicted in Figure 9d: Se atoms reaching the substrate attract the top-layer Ga atoms in the substrate and form bonds. Hence, the dark gap in Figure 8a between the initial GaSe TL and the substrate is referred to as a quasi-vdW (Q-vdW) gap, suggesting an interaction slightly stronger than true vdW bonding.

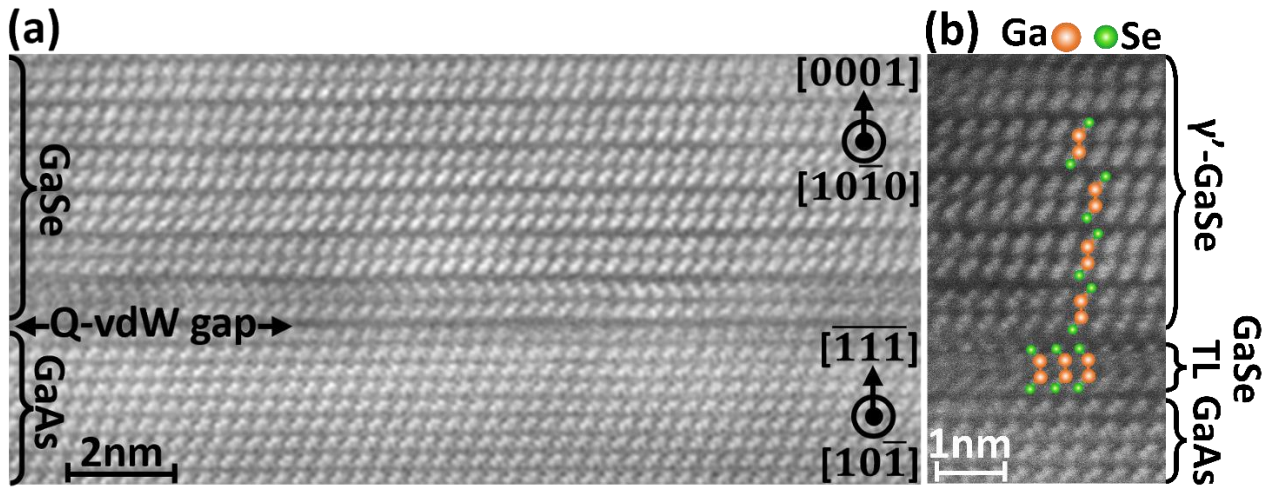

Figure 8. Cross-sectional ADF-STEM images of Sample #2 at (a) low magnification (low-pass filtered to reduce noise) and (b) high magnification.

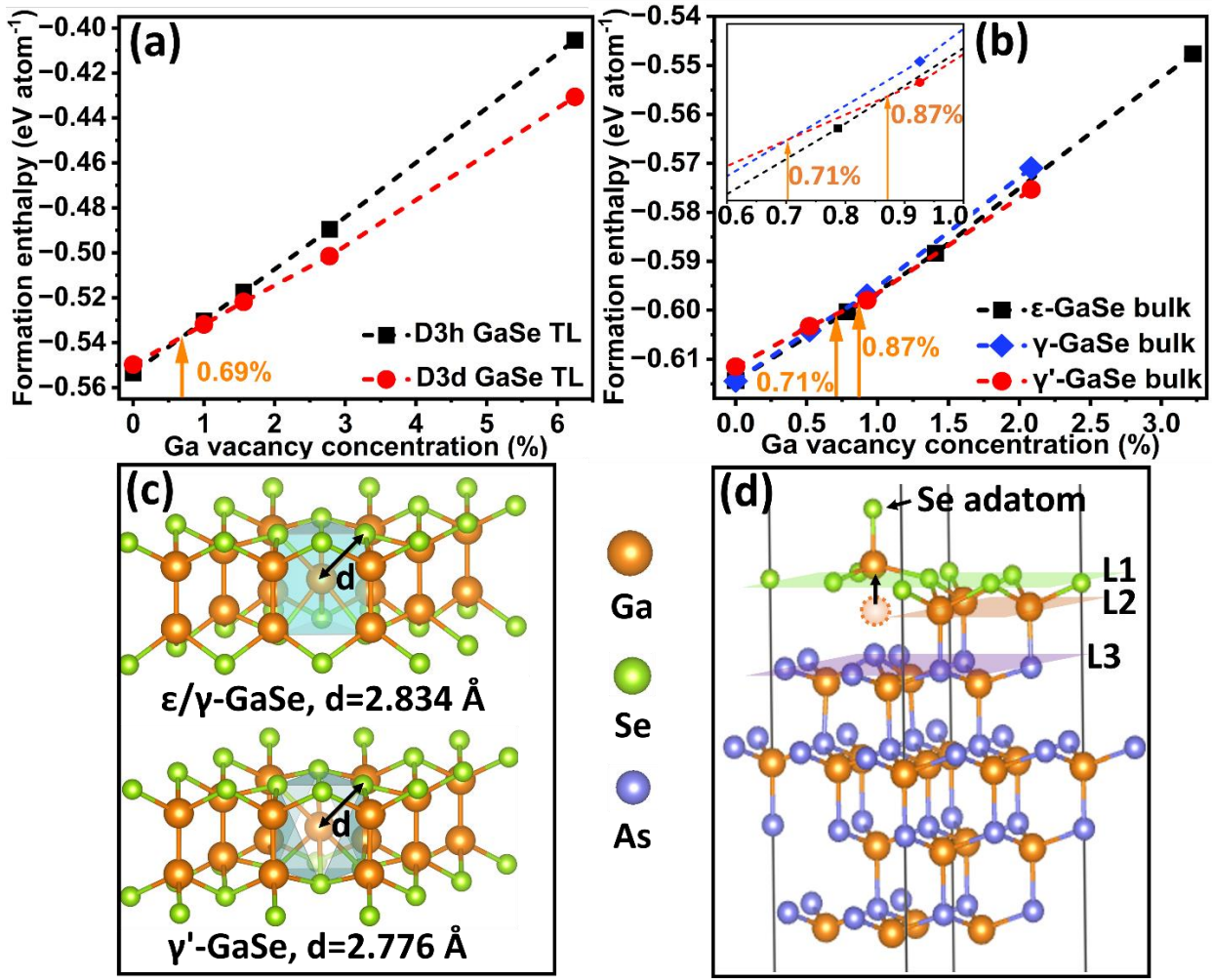

Figure 9. GaSe formation enthalpy as a function of Ga vacancy concentration in GaSe (a) TL and (b) bulk. Atomic models of the (c) γ'-GaSe TL structure (bottom panel) and the ε/γ-GaSe TL structure (top panel) after adding a Ga atom, "d" indicates the Ga-Se bond length. (d) An atomic model illustrating how Se adatoms that have just arrived at the surface of a Se-passivated GaAs(111)B substrate attract Ga atoms from the top layers of the substrate. L1, L2, and L3 indicate the Se, Ga, and As atomic planes on the surface of a Se-passivated GaAs(111)B substrate, respectively.

Another important finding of this study is that the GaSe growth on GaAs(111)B is significantly better than the growth on c-sapphire. We previously reported that GaSe films grown on c-sapphire substrates are misaligned with the substrate, rotated ~30 ° relative to the substrate orientation, and the GaSe layers are twisted with a wide distribution of angles, as depicted in Figure 10a.[32] In contrast, Figures 7c,f in this study demonstrate that the GaSe layer grown on GaAs(111)B has a strong, well-aligned epitaxial relationship with the substrate, and the GaSe layers are more ordered. This is reasonable as the sapphire substrate is a passivated substrate without surface dangling bonds, forming weak vdW bonds with the GaSe layer, leading to a more random initial orientation of the GaSe nuclei. Additionally, both our previous research (Figure 10b)[32] and the work of M. Shiffa *et al.*[34] have observed significant spherical 3D features in GaSe films deposited on c-sapphire. These features are attributed to the poor wettability of Ga adatoms on the sapphire surface, resulting in the balling up of the initial arriving Ga adatoms and the formation of 3D dot-like features composed of $Ga_2Se_3$. GaAs(111)B substrates address this issue due to the increased wettability of Ga on the Se-passivated GaAs(111)B surface, as evidenced by the large-area AFM scan (Figure 10c) where the 3D features are absent. Although a few very small features are observed, they are believed to be dust, as they only appear sporadically, in small quantities, and with uncertain shapes. Based on experimental and simulation results, the passivated GaAs(111) surface still interacts with the initial GaSe layer to some extent, which may assist in promoting the uniform nucleation/lateral expansion into a 2D GaSe layer rather than 3D features.

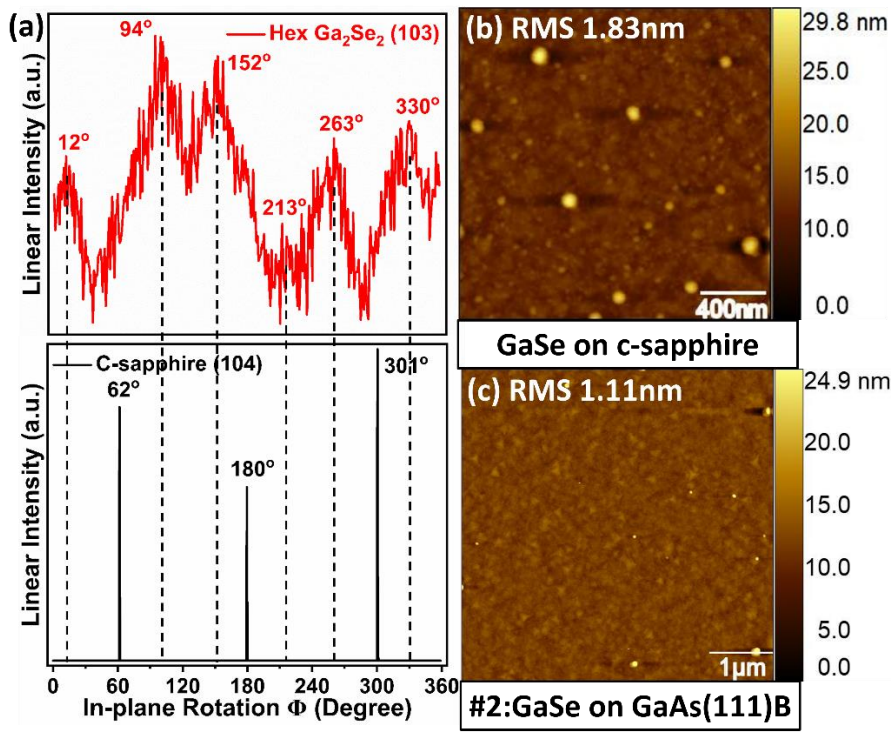

Figure 10. (a) In-plane φ scans of a GaSe film grown on c-sapphire substrate. The in-plane scans of the GaSe film and c-sapphire substrate were around the (103) and (104) planes, respectively. (b) 2 μm × 2 μm AFM image of a GaSe film grown on c-sapphire substrate. Part (a) and (b) are reproduced with permission.[32] Copyright 2023, American Vacuum Society. (c) 5 μm × 5 μm AFM image of Sample #2.

CONCLUSIONS

In summary, we have developed quasi-vdW epitaxial growth of single-crystal GaSe nanometer-thick films on GaAs(111)B substrates. Through a systematic study of growth parameters, we identified a set of optimal conditions: a Se:Ga flux ratio of 2.2, a growth temperature of 400 °C, and a growth rate of 0.07 Å s$^{-1}$. These parameters yielded GaSe crystal films with high structural quality: RMS surface roughness as low as 7 Å with a film thickness of 16 nm (*i.e.*, 20 TLs), and excellent alignment between the grown GaSe layer and the substrate. The GaSe layer consists of typical triangular domains that are primarily aligned with the major flat of the GaAs(111)B substrate but featuring numerous twin boundaries. The shape, orientation, and aggregation of GaSe domains are extremely sensitive to the supply of Se. A slight excess of Se leads to irregularly shaped and disordered nucleation and expansion of GaSe grains, reducing film smoothness but enhancing the crystallinity of GaSe. Higher growth temperatures are advantageous in suppressing screw dislocations and minimizing grain boundary defects caused by misoriented GaSe grains, though defects and droplets may arise due to GaSe evaporation at high temperatures. Increasing the growth rate helps reduce crystal defects.

It is noteworthy that the GaSe grown on GaAs(111)B is found to be the γ'-GaSe polymorph with a centrosymmetric TL, which has seldom been observed experimentally and is anticipated to function intriguingly in optics and optoelectronics. Finally, we show that a 3D GaAs(111)B substrate is more suitable for the epitaxial growth of 2D GaSe films compared to a passivated c-sapphire substrate. This study offers valuable insights for the epitaxial growth of 2D chalcogenide materials and provides a step towards fabricating high-quality hybrid 2D/3D heterostructures. In the future, further efforts to understand the quasi-vdW growth mechanism may prove to be crucial for the advancement of heterostructure semiconductor devices and integrated quantum communication systems.

METHODS

**MBE Growth.** GaSe films in this research were grown on epi-ready GaAs(111)B substrates using an R450 MBE reactor from DCA Instruments (instrument details at https://doi.org/10.60551/gqq8-yj90). The GaAs(111)B wafers, purchased from Wafer Tech, have a primary flat towards [01$\bar{1}$]. Prior to use, they were diced into 1 cm × 1 cm pieces and underwent sequential ultrasonic cleaning for 10 min each in acetone, isopropanol, and de-ionized water. After drying with a N$_2$ gun, the substrates were immediately transferred to the load lock chamber, where they were degassed for 2 hours at 200 °C in 5 × 10$^{-7}$ Torr to remove any residual contamination. Right before the growth of GaSe, the substrate was moved to the main chamber for deoxidation/Se-passivation treatment: the substrate was heated to 680 °C and annealed for 7 min, then cooled to the target growth temperature and held for 15 min to ensure temperature stability. The heating/cooling rate was maintained at 30 °C min$^{-1}$, and to inhibit substrate decomposition/evaporation at high temperatures, a Se flux of 1 × 10$^{14}$ atoms cm$^{-2}$ s$^{-1}$ was supplied when the substrate temperature exceeded 300 °C. The above parameters for processing GaAs substrates were determined by our previous work. For details, please refer to ref. 47. *In situ* RHEED was employed to monitor and confirm the removal of GaAs oxide and the subsequent GaSe growth. Ga and Se fluxes were independently provided by separate Knudsen effusion cells. The Se cell operated at relatively low temperatures (not exceeding 150 °C), generating uncracked Se molecules. Fluxes were measured by a quartz crystal microbalance (QCM) at the substrate position. The tooling factors of Se and Ga for QCM flux calibration were pre-determined through thickness measurements by X-ray reflectivity and cross-sectional SEM on calibration samples. The growth temperature was determined using a thermocouple positioned behind the substrate. The target thickness of GaSe samples for all growth campaigns in this work was 16 nm, *i.e.*, 20 TLs. After removal from the MBE chamber, all samples were stored individually in vacuum-sealed bags to minimize contamination and oxidation.

**Ex situ Characterization.** HRXRD 2θ/ω, ω, and in-plane φ scans were all performed on a Malvern PANalytical 4-Circle X'Pert 3 diffractometer equipped with a Cu-Kα$_1$ source. The 2θ/ω scans were utilized for identifying sample composition, phase, and crystallite size; ω scans were employed to evaluate crystal defects; in-plane φ scans were used to examine the symmetry, epitaxial quality, and in-plane ordering of the GaSe crystal films. On the other hand, the morphology and topology of the samples were observed using Bruker Dimension Icon AFM and Apreo SEM, and EDS mapping offered local elemental analysis. Raman spectroscopy, conducted on a Horiba LUCY equipped with a 532 nm laser, characterized the composition of the film and local area. Additionally, to image the intralayer atomic arrangements, we extracted electron-transparent cross sections utilizing an FEI Scios 2 dual-beam focused ion beam, and analyzed the cross sections through ADF-STEM using a dual spherical aberration-corrected FEI Titan[3] G2 60-300 STEM working at 300 kV, with a probe convergence angle of 21.3 mrad and collection angles of 42–244 mrad. All AFM, HRXRD, and Raman measurements were performed within 48 hours of sample removal from the MBE chamber. Furthermore, samples stored in vacuum-sealed bags showed no significant noticeable changes in surface morphology or crystal quality after six months, indicating that this preservation method is effective.

**DFT Calculation.** All the calculations were performed using the Vienna ab initio simulation package (VASP) code.[54] The effective core potentials were described by the projector augmented wave (PAW) potentials[55] with a cutoff energy of 600 eV. The Perdew-Burke-Ernzerhof (PBE) functionals[56] were used to describe the exchange-correlation interactions. The Monkhorst-Pack k-mesh was sampled with a density of 0.05 Å$^{-1}$. As for structure relaxation, BFGS quasi-Newton algorithm was used, and the thresholds of convergence used 10$^{-5}$ eV as a break condition for the electronic self-consistence loop, and the Hellmann-Feynman force on each atom was less than 0.01 eV Å$^{-1}$. To account for the vdW effect, the non-local vdW-DF-optB88 exchange correlation functional[57] was applied to describe the dispersion interactions within the interface. A vacuum layer of 18 Å thickness was used for all 2D structures to eliminate interactions between adjacent supercells due to periodic boundary conditions (PBCs). The valence electron configurations are $4s^24p^1$ for Ga, $4s^24p^3$ for As, and $4s^24p^4$ for Se. The in-plane lattice parameters of ε- and γ-GaSe are both 3.827 Å, while that of γ'-GaSe is 3.839 Å. The interlayer distances in ε-, γ-, and γ'-GaSe are 3.209 Å, 3.206 Å, and 3.221 Å, respectively. These data are comparable with references.[29,34] To investigate the energetic information of GaSe polymorphs, we calculated the formation enthalpy through DFT method follow the formular:

$$\Delta H = \frac{E_{total} - n_{Ga}E_{Ga_8}^{bulk} - n_{Se}E_{Se_{32}}^{bulk}}{n_{total}}$$

where $E_{total}$ is the total energy of system, $E_{Ga_8}^{bulk}$ and $E_{Se_{32}}^{bulk}$ are the chemical potentials of each atomic species in the most stable form. $n_{Ga}$ and $n_{Se}$ are the number of Ga and Se atoms, respectively, and $n_{total}$ is the total number of atoms in the supercell.

ASSOCIATED CONTENT

This manuscript has been previously submitted to a pre-print server as: Yu, M.; Iddawela, S. A.; Wang, J.; Hilse, M.; Thompson, J. L.; Hickey, D. R.; Sinnott, S. B.; Law, S. Quasi-van der Waals Epitaxial Growth of γ'-GaSe Thin Films on GaAs(111)B Substrates. 2024, 12265. arXiv. arXiv:2403.12265 (accessed May 23, 2024).

AUTHOR INFORMATION

Corresponding Author

Stephanie Law - Department of Materials Science and Engineering, Materials Research Institute, 2D Crystal Consortium Materials Innovation Platform, Penn State Institute of Energy and the Environment, The Pennsylvania State University, University Park, PA 16802, USA; Email: sal6149@psu.edu

Author Contributions

M.Y. conducted the thin film synthesis, sample characterization, data analysis, and manuscript writing. S.A.I. and J.L.T. performed the STEM imaging, D.R.H. participated in coordinating this work. J.W. performed DFT calculations and S.B.S. guided this work. M.H. provided help in film synthesis and paper editing. S.L. supervised this project. The manuscript was written through contributions of all authors. All authors have given approval to the final version of the manuscript.

Funding Sources

NSF cooperative agreements DMR-203931. Coherent/II-VI Foundation; Startup funds from the Penn State Eberly College of Science, Department of Chemistry, College of Earth and Mineral Sciences, Department of Materials Science and Engineering, and Materials Research Institute. NSF through the Penn State University Materials Research Science and Engineering Center DMR-2011839. Funding from the Basic Office of Science of the Department of Energy under Award DE-SC0018025.

Notes

The authors declare no competing financial interest.

ABBREVIATIONS

2D, two-dimensional; 3D, three-dimensional; vdW, van der Waals; MBE, molecular beam epitaxy; TL, tetralayer; ADF-STEM, annular dark field-scanning transmission electron microscopy; DFT, density functional theory; c-sapphire, c-plane sapphire; HRXRD, high-resolution X-ray diffraction; FWHM, full-width at half maximum; AFM, atomic force microscopy; RMS, root mean square; SEM, scanning electron microscopy; EDS, electron dispersive spectroscopy; RHEED, reflection high-energy electron diffraction; Q-vdW, quasi-van der Waals; QCM, quartz crystal microbalance; VASP, Vienna ab initio simulation package; PAW, projector augmented wave; PBE, Perdew-Burke-Ernzerhof; PBCs, periodic boundary conditions.

ACKNOWLEDGMENT

This research was conducted at the Pennsylvania State University Two-Dimensional Crystal Consortium – Materials Innovation Platform which is supported by NSF cooperative agreement DMR-2039351. M. Y. and S. L. acknowledge funding from the Coherent/II-VI Foundation. S.A.I., J.L.T., and D.R.H. acknowledge support through startup funds from the Penn State Eberly College of Science, Department of Chemistry, College of Earth and Mineral Sciences, Department of Materials Science and Engineering, and Materials Research Institute. J.L.T. also acknowledges research support by NSF through the Penn State University Materials Research Science and Engineering Center DMR-2011839. J.W. and S.B.S. acknowledge funding from the Basic Office of Science of the Department of Energy under Award DE-SC0018025. The authors appreciate the use of the Penn State Materials Characterization Lab. They also acknowledge the use of the computational facilities associated with the Institute for Computational and Data Science at Penn State University.

DATA AVAILABILITY: The data sets that support the findings of this study are openly available at https://doi.org/10.26207/pgtf-5570

Supporting Information Available:

[1]Table summarizing growth parameters of GaSe sample #1 – #15 grown on GaAs(111)B substrates. [2]Profile scan along the white solid line in Figure 3b. [3]Optical microscopy image, SEM image, Raman spectra, and EDS mapping of Sample #8. [4]AFM images of Sample #9 – #11. [5]HRXRD 2θ/ω and ω patterns of Sample #9 – #15. [6]GaSe formation enthalpy as a function of Se vacancy concentration in GaSe TL.

BRIEFS

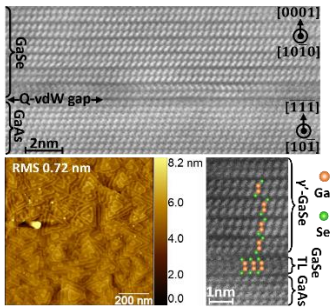

Table of Content (TOC) Figure: MBE-grown GaSe thin films on GaAs(111)B substrate exhibit smooth surfaces and a rare unit-layer centrosymmetric space group.

SYNOPSIS

We developed a technique for the direct growth of GaSe single crystal nanometer-thick films on GaAs substrates using molecular beam epitaxy, forming a 2D/3D heterostructure with a well-defined interface. The obtained GaSe thin films not only exhibit extremely low surface roughness but also display a rare pure γ'-polymorph with a centrosymmetric unit layer. Through density-functional theory calculations, we identified that the energetically unfavorable γ'-GaSe arises from the involvement of Ga vacancies during growth. This γ'-polymorph is anticipated to show distinct properties such as second harmonic generation and enhanced optoelectronic characteristics.

# Quasi-van der Waals Epitaxial Growth of γ'-GaSe Nanometer-Thick Films on GaAs(111)B Substrates

Mingyu Yu,[†] Sahani A. Iddawela,[‡] Jiayang Wang,[§] Maria Hilse,[§,‖,⊥] Jessica L. Thompson,[‡] Danielle Reifsnyder Hickey,[‡,§,‖] Susan B. Sinnott,[‡,§,‖,#,¶] Stephanie Law[*,§,‖,⊥,¶]

AUTHOR ADDRESS

[†]Department of Materials Science and Engineering, University of Delaware, 201 Dupont Hall, 127 The Green, Newark, DE 19716 USA

[‡]Department of Chemistry, The Pennsylvania State University, University Park, PA 16802 USA

[§]Department of Materials Science and Engineering, The Pennsylvania State University, University Park, PA 16802 USA

[‖]Materials Research Institute, The Pennsylvania State University, University Park, PA 16802, USA

[⊥]2D Crystal Consortium Materials Innovation Platform, The Pennsylvania State University, University Park, PA 16802, USA

[#]Institute for Computational and Data Science, The Pennsylvania State University, University Park, PA 16802, USA

[¶]Penn State Institute of Energy and the Environment, The Pennsylvania State University, University Park, PA 16802 USA

Corresponding Email: sal6149@psu

Table S1. Growth parameters of GaSe Sample #9 – #15 grown on GaAs(111)B substrates. The growth parameters of Sample #1 – #8 are summarized in Table 1 of the main text. According to calibration results, Ga fluxes of 1.3 (± 0.1) × $10^{13}$, 2.3 × $10^{13}$, 0.9 × $10^{13}$, 0.7 × $10^{13}$, and 0.5 × $10^{13}$ atoms $cm^{-2}$ $s^{-1}$ can lead to GaSe growth rates of 0.07, 0.13, 0.05, 0.04, and 0.03 Å $s^{-1}$, respectively, when there is no re-evaporation.

| Sample | Ga flux [$10^{13}$ atoms $cm^{-2}$ $s^{-1}$] | Se:Ga | Growth temperature [°C] | Growth time [s] |
|---|---|---|---|---|
| #9, #13 | 1.3 | 2.2 | 400 | 2400 |
| #10, #14 | 0.9 | 2.2 | 400 | 3360 |
| #11, #15 | 0.5 | 2.2 | 400 | 5600 |
| #12 | 0.7 | 2.2 | 400 | 4200 |

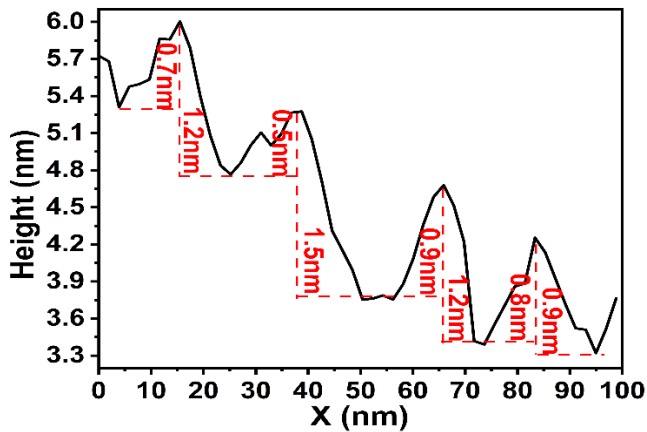

Figure S1. A profile scan along the white solid line in Figure 3b in the main text. It shows an average step height of about 0.8 nm. The height difference between adjacent layers is close to the GaSe tetralayer (TL) thickness, indicating the layer-by-layer mode of molecular beam epitaxy growth.

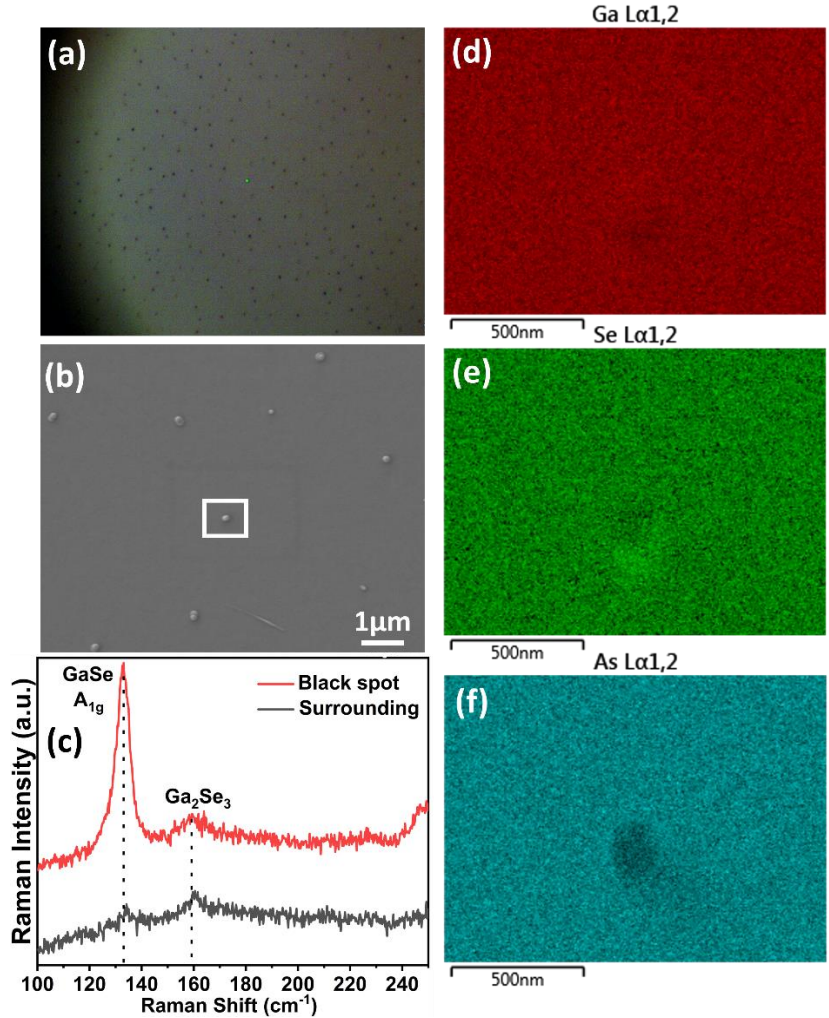

Figure S2. (a) Optical microscopy image, (b) scanning electron microscopy image, (c) Raman spectra of Sample #8. The "black spot" in part (c) refer to the black spots distributed on part (a). (d-f) Energy dispersive spectroscopy (EDS) mapping for (d) Ga, (e) Se, and (f) As elements. EDS mapping was conducted in the area marked by the white rectangle in part (d). Sample #8 was grown at 420 °C, a high temperature that causes some evaporation of GaSe, leading to droplets on the film surface, which appear as three dimensional (3D) features in parts (a) and (b). To analyze

the components of these features, part (c) compares a Raman scan around the black spot marked in part (a) with a Raman scan of the surrounding area. Both curves show signals from GaSe $A_{1g}$ mode and $Ga_2Se_3$. The existence of $Ga_2Se_3$ phase is attributed to the degradation of GaSe in the air and light.[1] However, notably, the black spot shows a more intensive GaSe signal than the surrounding area, demonstrating that more GaSe accumulated on these black spots. Furthermore, the EDS maps (d-f) show that the proportion of As content in the 3D feature is obviously lower than that of the surrounding area, while the proportion of Se content is higher than that of the surrounding area. Both Raman spectra and EDS mapping suggest that these droplets are composed of GaSe.

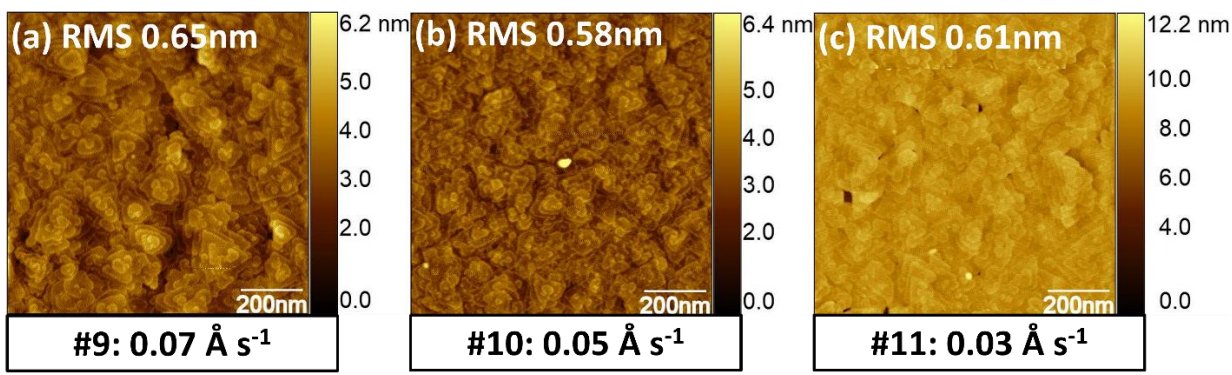

Figure S3. Atomic force microscopy (AFM) images of Sample (a) #9, (b) #10, and (c) #11. "RMS" stands for root-mean-square roughness. The three samples used different growth rates while other growth conditions were the same. From the AFM images, different growth rates did not lead to significant differences in surface morphology or surface roughness.

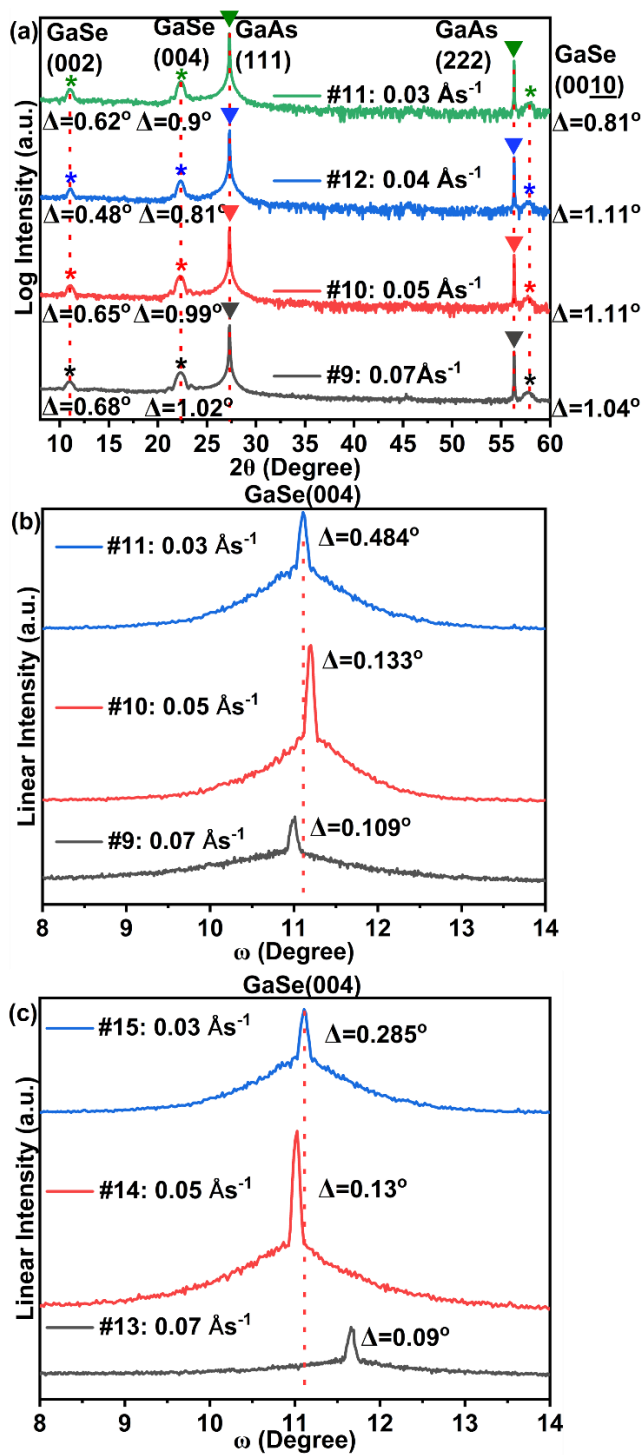

Figure S4. High resolution X-ray diffraction patterns. a) 2θ/ω scans of Sample #9 – #12, "*" and "▼" symbols mark the peaks of GaSe and GaAs, respectively; "Δ" indicates the value of full width at half maximum (FWHM). The four samples used different growth rates while other growth

conditions were the same. From 2θ/ω scans, we conclude that different growth rates did not lead to significant differences in film composition or crystallite size. (b-c) ω scans of Sample (b) #9 – #11 and (c) #13 – #15. ω scans were conducted around the GaSe(004) plane that was the strongest GaSe diffraction peak in 2θ/ω scans. The two samples in each group: 1) #9 and #13, 2) #10 and #14, 3) #11 and #15, used the same growth conditions, while #9 – #11 and #13 – #15 are two growth campaigns. We made two samples for each growth rate to confirm the accuracy and authenticity of the data. In each campaign, the FWHM value decreases with increasing growth rate, indicating that higher growth rate in beneficial to reduce crystal defects and improve crystallinity. The average of the FWHM values is shown in Figure 6 of the main text.

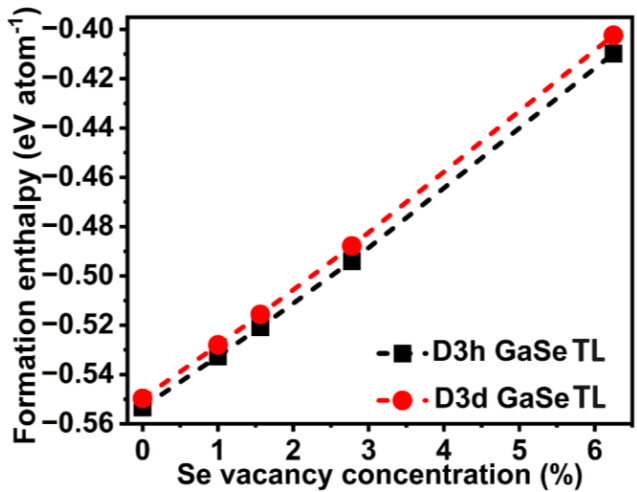

Figure S5. GaSe formation enthalpy as a function of Se vacancy concentration in GaSe TL. Using density-functional theory, we calculated the formation enthalpy of GaSe TL when Se vacancies are introduced into the GaSe crystal structure. Regardless of changes in Se vacancy concentration, the formation enthalpy of $D_{3h}$ GaSe TL (the common polymorph in GaSe crystals) is always lower than that of $D_{3d}$ GaSe TL (a rare polymorph observed in γ'-GaSe). This indicates that Se vacancies cannot lead to the formation of γ'-GaSe.